\documentclass[3p,times,twocolumn]{elsarticle}

\usepackage{ecrc}

\volume{00}

\firstpage{1}

\journalname{Nuclear and Particle Physics Proceedings}

\runauth{Nausheen R. Shah}

\jid{nppp}

\jnltitlelogo{Nuclear and Particle Physics Proceedings}

\usepackage{amssymb}

\usepackage[figuresright]{rotating}

\begin{document}

\begin{frontmatter}


 \title{The Higgs and WIMP DM Lamp Posts for New Weak Scale Physics:\\
 EFT Perspectives and the NMSSM\tnoteref{label1}}
\tnotetext[label1]{Based on talk presented at the Seventh Workshop on Theory, Phenomenology
and Experiments in Flavour Physics and the future of BSM physics, Capri, June 2018.}
 \author{Nausheen R. Shah
 }
 \ead{nausheen.shah@wayne.edu}
 \ead[url]{https://clasprofiles.wayne.edu/profile/fz9187}
 \address{Department of Physics \& Astronomy, Wayne State University, Detroit, MI 48201, USA
 }

\dochead{\small WSU-HEP-1807}





\begin{abstract}
I will show, via effective field theory~(EFT) techniques, that obtaining an observationally consistent relic density while evading stringent direct detection limits and maintaining $h_{125}$ phenomenology in an extended Higgs sector can be easily achieved. I will then map such an EFT to the low energy limit of the NMSSM with the Higgsinos integrated out. Both the singlino and the singlet-like CP-odd and even scalars in the NMSSM may play a relevant role in such a scenario, while being difficult to probe via conventional searches. The singlet sector of the general NMSSM can be mapped on to a 2HDM+S, and I will discuss prospects of probing this at the LHC using signatures such as mono-Higgs and mono-Z. This proceeding is mostly based on Refs.~\cite{Baum:2017enm} and~\cite{Baum:2018zhf}.     
\end{abstract}


\end{frontmatter}


\section{Introduction}
\label{sec:intro}

In recent years the WIMP paradigm has become increasingly unpopular due to the stringent limits being placed on the spin-independent direct detection~(SIDD) scattering cross section of dark matter~(DM) with nuclei from experiments such as XENON1T~\cite{Aprile:2017iyp}. Analogously, the null direct search results at the LHC, and the so far very Standard Model~(SM) like nature of $h_{125}$ has lead to a general pessimism in our field regarding the existence of an extended Higgs sector. However, if certain relationships between model parameters are fulfilled, it is easy to evade both $h_{125}$ considerations and dark matter scattering limits, while obtaining an observationally consistent dark matter relic density. Ultimately we are in search of the ultraviolet~(UV) completion of the SM, which we expect to be based on symmetries at the UV scale. In such a case it is not difficult to imagine that low-scale physics may show up as having certain ``fine-tuned" relations. Keeping an agnostic attitude towards such symmetries, one can nevertheless investigate the requirements for the parameters of a model such that our current lack of experimental signals is not due to the absence of new physics~(NP) at the weak-scale, but rather because certain relationships exits making NP challenging to discover with conventional means.

In Sec.~\ref{sec:EFT} I  present a minimal scenario for obtaining a consistent relic density and a suppressed SIDD cross-section from an extended Higgs sector and an additional SM singlet playing the role of DM, using the language of EFT. I  show that these requirements can be easily fulfilled without any unnatural hierarchies in the model parameters or physical properties. Sec.~\ref{sec:LHC}  outlines  LHC search strategies and their prospects which may be used to probe such a scenario. I  then show in Sec.~\ref{sec:NMSSM} how such a model is mapped onto the Next to Minimal Supersymmetric Standard Model~(NMSSM).  I  summarize in Sec.~\ref{sec:sum}.

\section{EFT for Higgs Couplings to Dark Matter}\label{sec:EFT}

\begin{figure}[t]
\hspace*{7mm}
\includegraphics*[width=.95\linewidth]{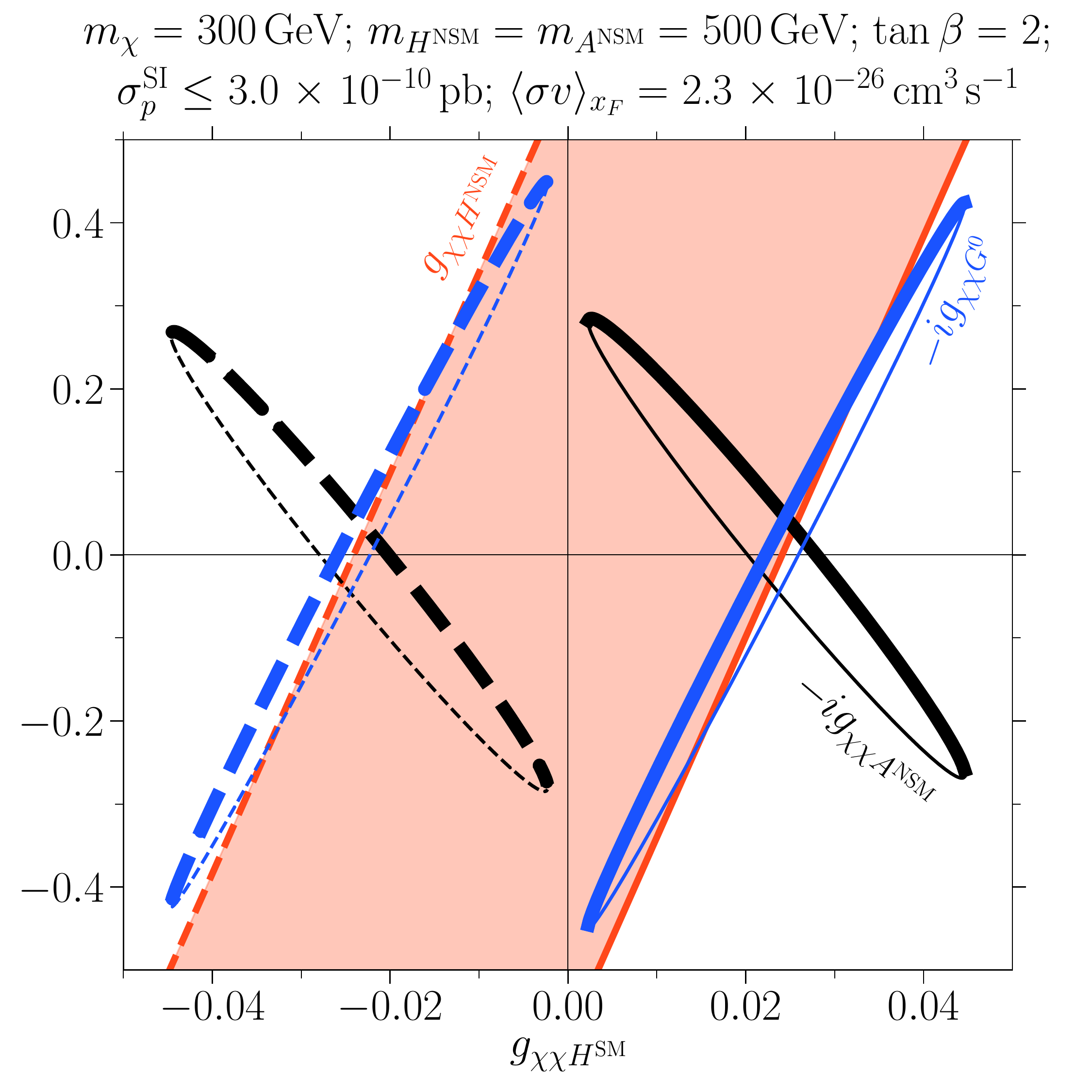} 
  \caption{\label{eftCoup} 
  EFT parameters and couplings of DM to the CP-even and CP-odd Higgs bosons required to obtain the correct thermal relic density while concurrently satisfying SIDD constraints, for $\tan\beta = 2$, $m_\chi = 300\,$GeV, $m_{H^{\rm NSM}} = m_{A^{\rm NSM}} = 500\,$GeV, and decoupled singlet states.  The orange shaded-region bounded by solid and dashed lines represents the CP-even Higgs bosons couplings consistent with the SIDD bounds, while the blue and black ellipses denote the couplings of DM to the (neutral) Goldstone mode and the heavy CP-odd Higgs boson yielding $\Omega h^2\sim0.12$ for CP-even Higgs couplings denoted by the corresponding solid or dashed lines, with thick and thin lines denoting two different solutions~\cite{Baum:2017enm}.}
\end{figure}

\begin{figure}[t]
\hspace*{7mm}
\includegraphics*[width=.95\linewidth]{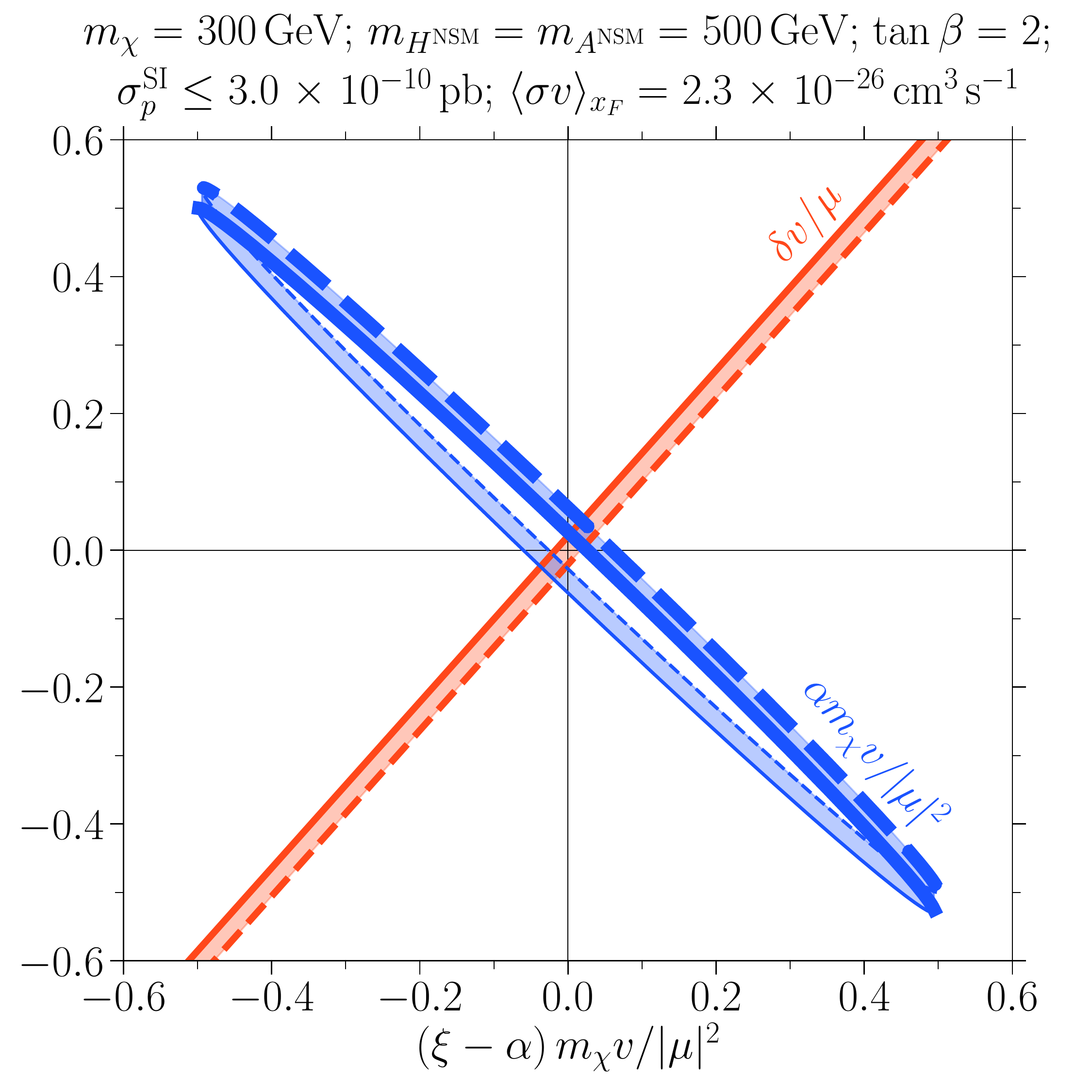} 
  \caption{\label{Pars} Values of the EFT parameters consistent with the couplings shown in Fig.~\ref{eftCoup}. Dashed and solid lines, as well as the shaded areas shown in this panel are in one-to-one correspondence with similar 
		lines and areas shown in the left panel~\cite{Baum:2017enm}.
}
\end{figure}

I will discuss a model with a SM singlet Majorana fermion DM $\chi$, which has no renormalizable interactions with SM particles. In order to couple DM to the SM, we consider the Higgs sector of a Type II Two Higgs doublet model~(2HDM): $H_u$ and $H_d$. 

Assuming, as usual, that both Higgs doublets acquire vacuum expectation values~(vevs), $\langle H_d \rangle = v_d$, $\langle H_u \rangle = v_u$, with $(v_d^2+v_u^2) = (174\mathrm{~GeV})^2$ and $\tan\beta = v_u/v_d$, I define the Higgs basis~\cite{Georgi:1978ri, Donoghue:1978cj, gunion2008higgs, Lavoura:1994fv, Botella:1994cs, Branco99, Gunion:2002zf} such that the mass eigenstate ~($H^{\rm SM})$ associated with the observed 125 GeV Higgs boson has completely standard model couplings, i.e the SM vev is acquired by the field corresponding to the neutral component of $H^{\rm SM}$, hence $\langle H^{\rm SM}\rangle = \sqrt{2} v$ and $\langle H^{\rm NSM}\rangle = 0$: 
\begin{eqnarray}
	H^{\rm SM} &= \sqrt{2} {\rm Re} \left( \sin\beta H_u^0 + \cos\beta H_d^0 \right), \label{eq:Hbasis1}
	\\ G^0 &= \sqrt{2} {\rm Im} \left( \sin\beta H_u^0 - \cos\beta H_d^0 \right),
	\\ H^{\rm NSM} &= \sqrt{2} {\rm Re} \left( \cos\beta H_u^0 - \sin\beta H_d^0 \right),
	\\ A^{\rm NSM} &= \sqrt{2} {\rm Im} \left( \cos\beta H_u^0 + \sin\beta H_d^0 \right),
	\label{eq:Hbasis-1}
\end{eqnarray} 

In addition, we impose that there are no explicit mass terms or scales and hence the Lagrangian is scale invariant. The absence of explicit scale dependence could be originating from a $Z_3$ symmetry. In such a situation, a natural way to generate the mass $m_\chi$ and the scale of NP $\mu$ is via the vev of a singlet $S= \langle S \rangle+\frac{1}{\sqrt{2}} \left(H^S + i A^S \right)$. Hence, without loss of generality we can define $m_\chi = 2 \kappa\langle S \rangle$ and $\mu = \lambda \langle S \rangle$, where $\kappa$ and $\lambda$ are dimensionless parameters.

Assuming that  $d > 4$ terms 
originate from a theory where a heavier $SU(2)$-doublet Dirac fermion with mass $\mu$ has been integrated out, we can write all the allowed $d=6$ operators which would arise from integrating out such a field, describing the interactions of a Majorana fermion $\chi$ with the two Higgs doublets $H_u$, $H_d$. Ignoring the charged gauge boson interactions, we get~\cite{Baum:2017enm} 
\begin{eqnarray} \label{eq:EFTmu}
	&&\mathcal{L} =~ - \delta \frac{\chi\chi}{\mu} \left( H_u \!\cdot\! H_d \right)\left( 1 - \frac{\lambda \hat{S}}{\mu} \right)  \nonumber \\	
	&&- \kappa S \chi\chi \left(1 + \xi \frac{H_d^\dagger H_d + H_u^\dagger H_u }{| \mu |^2} \right) + {\rm h.c.}\nonumber \\
	&&+ \frac{\alpha}{|\mu|^2}\left\{\chi^\dagger H_u^\dagger \bar{\sigma}^\mu \left[ i \partial_\mu - \frac{g_1}{s_W} (T_3 - Q s^2_W) Z_\mu \right] (\chi H_u) \right. \nonumber \\
	&& + \left. \chi^\dagger H_d^\dagger \bar{\sigma}^\mu \left[ i \partial_\mu - \frac{g_1}{s_W} (T_3 - Q s^2_W) Z_\mu \right] (\chi H_d) \right\} , \nonumber \\
\end{eqnarray}
where $S = \mu/\lambda + \hat{S}$, $Q$ and $T_3$ are the charge and weak isospin operators, $s_W \equiv \sin \theta_W$ with the weak mixing angle $\theta_W$, and $g_1 = e/\cos \theta_W$ is the hypercharge coupling. 

From Eqs.~(\ref{eq:Hbasis1}) and~(\ref{eq:EFTmu}), the coupling of the DM particles to the SM-like Higgs is
\begin{equation} \label{eq:gxxHSM1}
	g_{\chi\chi h} \simeq g_{\chi\chi H^{\rm SM}} = \frac{\sqrt{2} v}{\mu} \left[\delta \sin 2\beta - \frac{(\xi -\alpha)m_\chi}{ \mu^*} \right] .
\end{equation} 

The SIDD scattering cross-section is mediated by the $t$-channel exchange of the CP-even scalars. Generically, the SM-like Higgs has the dominant contribution. 
However,
we see that we can obtain a blind spot for the cancellation of the coupling of $H^{\rm SM}$ to pairs of DM occurs for
\begin{equation}\label{eq:bs}
	\sin 2 \beta = \frac{(\xi-\alpha) m_\chi}{\mu^* \delta}\;.
\end{equation}
In proximity to such a blind spot, there can be further suppression of the SIDD due to interference with the other CP-even Higgs bosons. 

Note that the couplings to the other Higgs states, including the Goldstone modes, are not necessarily suppressed. In fact there can be significant couplings of the extended Higgs sector states to DM, mediating the annihilation cross-section needed for obtaining an observationally consistent relic density. It turns out that $\Omega h^2 \sim 0.12$ can be easily achieved by the $s$-channel annihilation of the DM particles into $t\bar{t}$ mediated by  $G^0$ for the region of masses we are investigating. 

An example for couplings of the Higgs sector states to DM where the SIDD is suppressed while simultaneously obtaining a consistent relic density is shown in Fig.~\ref{eftCoup}. The corresponding EFT parameters are shown in Fig.~\ref{Pars}. The mass spectrum is fixed to labeled values. Changing these values would change the precise numerical values shown, but the qualitative behavior remains the same. I stress that neither the couplings nor the EFT parameters have any extreme values. While specific relationships between parameters have to be fulfilled, the resulting scenario does not appear to be particularly difficult to achieve.   

I have discussed how to obtain a WIMP with thermally produced relic density whose direct detection is suppressed, in conjunction with a Higgs sector which is aligned so that $h_{125}$ phenomenology is completely SM like. 
The couplings between the Higgs states don't play a role in the scenario discussed so far.  However, the details of the Higgs scalar potential may be relevant for the discovery prospects for such a scenario, and some possibilities will be discussed in the next sections.

\section{Higgs Phenomenology \& LHC prospects}\label{sec:LHC}

The scalar potential for a 2HDM + S is given in Refs.~\cite{Carena:2015moc, Baum:2018zhf}. The generic potential is described by 27 arbitrary parameters and at first glance appears difficult to analyze. However the 125 GeV Higgs mass and its SM-like couplings enable us to constrain these significantly. In particular, we show that most of the relevant phenomenology can be parameterized in terms of mostly the physical parameters like masses and mixing angles~\cite{Baum:2018zhf}. 

I highlight first a few conditions that alignment imposes on the phenomenology. The most important thing is that alignment forbids the coupling of the NSM or S like CP-even Higgs bosons from coupling to pairs of $h_{125}$ or vector bosons~($W$ or $Z$). Additionally the CP-odd state couplings to $h_{125}$ and $Z$ are also forbidden. Instead, there can be interesting {\it Higgs cascade decays} of the heavy Higgs bosons to final states involving only {\it one} $h_{125}$ {\it or} a $Z$ such as $(H^{\rm{NSM}} \to H^{\rm{S}} H^{\rm{SM}})$ or $(A^{\rm{NSM}} \to H^{\rm{S}} Z)$. The singlets couple only to DM or to the SM particles via their mixing with the other states. Hence depending on the mixing angles and the arbitrary coupling to the DM, such decays could result in $h_{125}$ or $Z$ plus visible or invisible signatures.  

\begin{figure}[tbh]
\hspace*{1mm}
\includegraphics*[width=.95\linewidth]{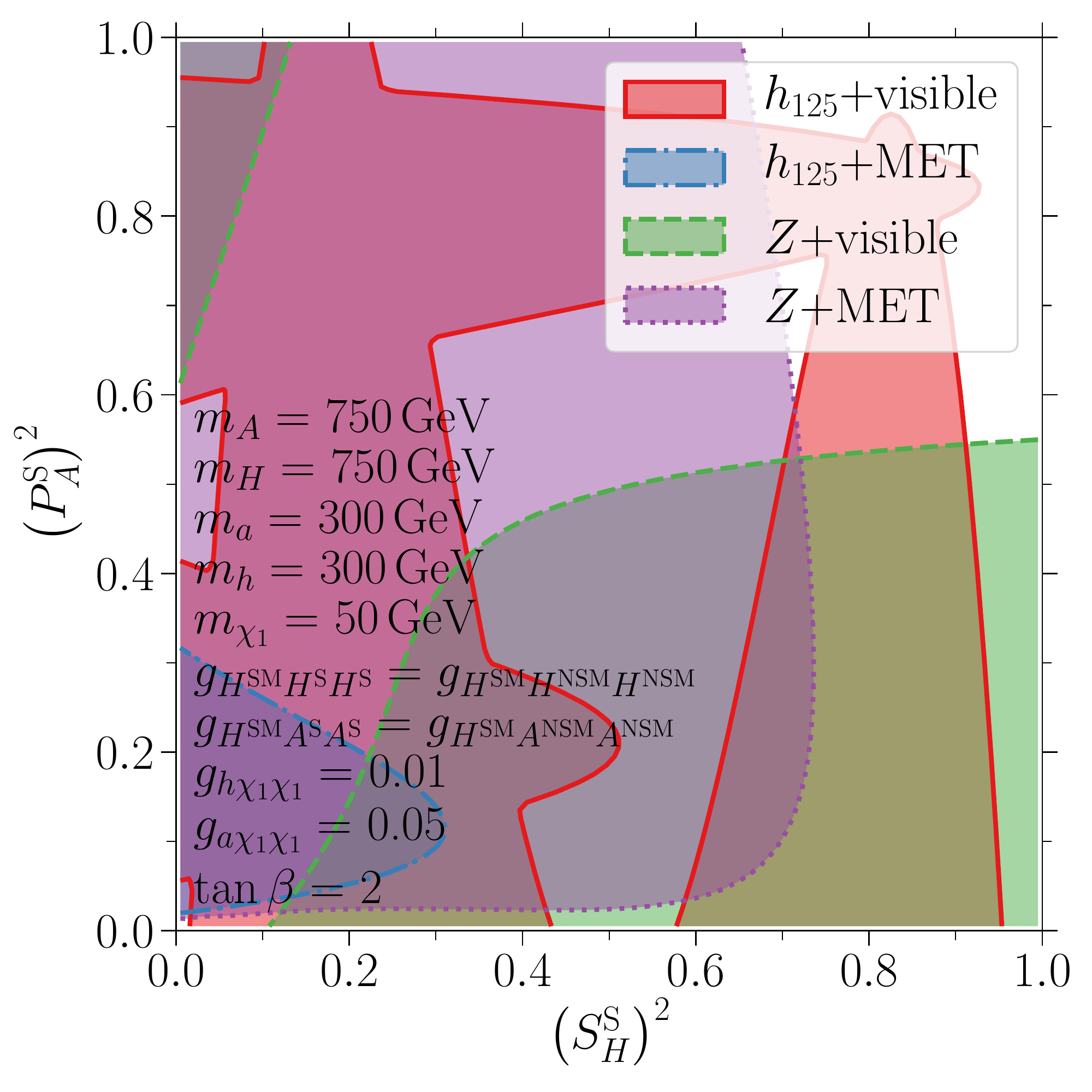} 
  \caption{\label{reach_mix} 
  Regions of 2HDM+S parameter space within the future reach of the different Higgs Cascade search modes as indicated in the legend at the LHC with $L = 3000\,{\rm fb}^{-1}$ of data. The colored regions show the accessible regions via the various signatures in the plane of the singlet fraction of the parent Higgs bosons $(S_H^{\rm S})^2$ vs $(P_A^{\rm S})^2$~\cite{Baum:2018zhf}. }
\end{figure}

\begin{figure}[tbh]
\hspace*{1mm}
\includegraphics*[width=.95\linewidth]{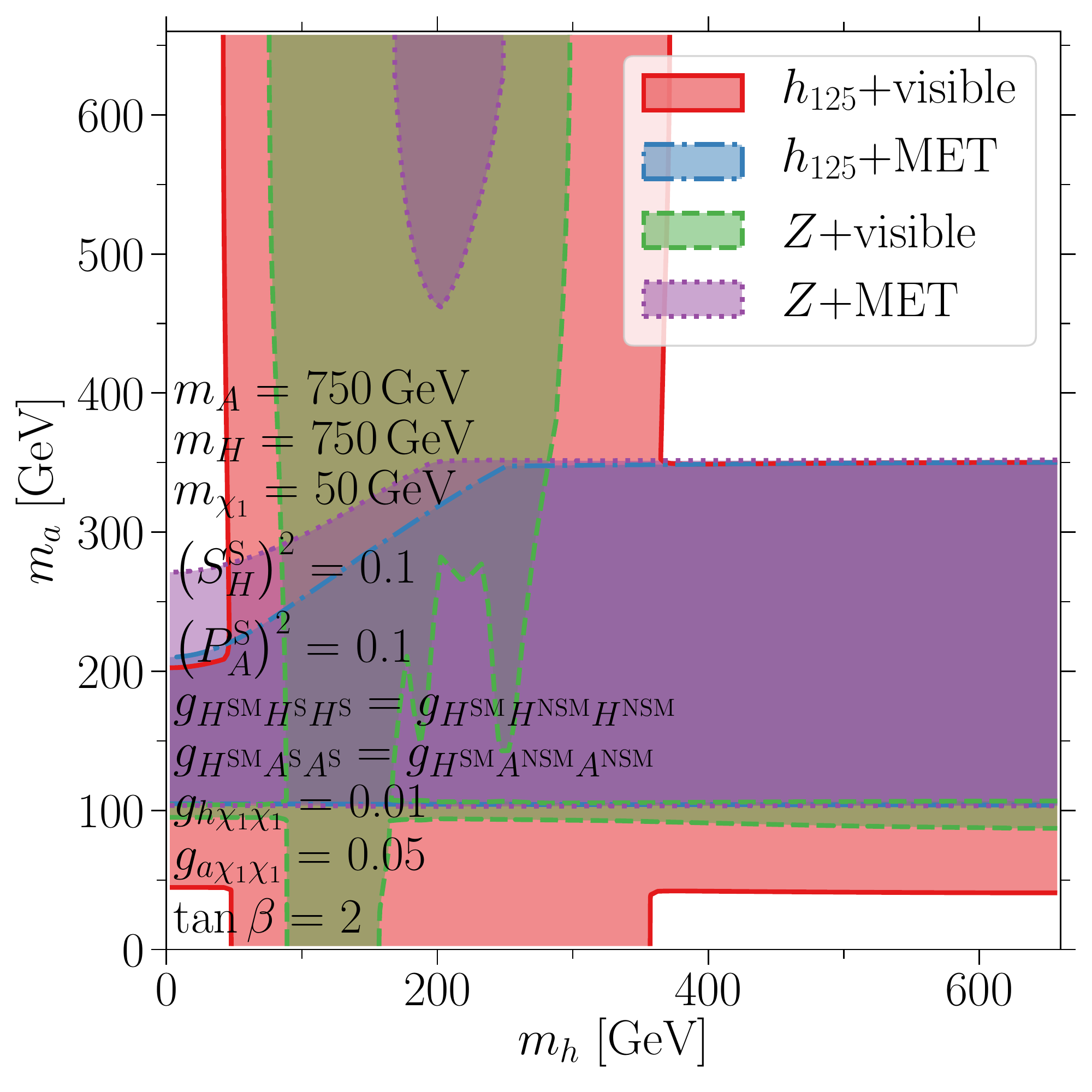} 
  \caption{\label{reach_mass} 
Same as Fig.~\ref{reach_mix}, but with the reach shown in the plane of the masses of the daughter Higgs bosons produced in the Higgs Cascades, $m_h$ vs $m_a$. The remaining parameters are fixed to the values indicated in the labels~\cite{Baum:2018zhf}.}
\end{figure}

We collected all the current search results and projections available for the relevant decays, as well as performed detailed collider simulations where needed, to obtain the projection for the reach at the LHC with 3000 fb$^{-1}$ of data. I present an example of the reach we obtain for exemplary scenarios in Figs.~\ref{reach_mix} and~\ref{reach_mass}. As can be seen, combining the different searches for the various Higgs cascade decay modes provides coverage of most of the parameter space at low values of $\tan\beta$, a region which is generally challenging to probe~\cite{Gori:2016zto}. 

\section{NMSSM Interpretation}\label{sec:NMSSM}

The Next to Minimal Supersymmetric Standard Model~(NMSSM) is a well motivated extension of the SM. 
The NMSSM has a 2HDM + Singlet~(S) scalar sector, analogous to the Higgs sector assumed in the previous section. What makes the NMSSM particularly interesting is that a Higgs boson with a mass of 125 GeV and SM-like couplings is easily and naturally obtained~\cite{Carena:2015moc}. The NMSSM provides two SM Majorana singlets which may play the role of DM, the singlino~(superpartner of the Singlet Higgs) and the bino~(superpartner of the hypercharge gauge boson). Due to SUSY relations, phenomenological considerations of the $h_{125}$ correlate the masses of all the scalars as well as that of the Singlinos and the Higgsinos, leading to a consistent scenario where the entire Higgs sector as well as the DM candidates can be $\sim\mathcal{O}(\rm {few~ } 100)$ GeV.   

The DM-Higgs EFT discussed in Sec.~\ref{sec:EFT} can be trivially mapped to the NMSSM in the following two regions~\cite{Baum:2017enm}: 
\begin{itemize}
\item For Singlino DM, we can map the couplings in Eq.~(\ref{eq:EFTmu}) directly to those in the NMSSM via
\begin{equation} \label{eq:pmapS}
	\delta = -\alpha \to -\lambda^2~,  \lambda \to \lambda~,  \kappa \to \kappa~,  \xi \to 0~. 
\end{equation}
The mapping above leads to the blind spot condition [cf. Eq.~(\ref{eq:bs})]
\begin{equation}\label{eq:bsSin}
	\sin 2\beta = m_\chi/\mu~.
\end{equation}
\item In contrast to the singlino, the bino couples to different combinations of the Higgs doublets and the singlet. Such interactions would be obtained by writing down the EFT for the Higgs doublets and the singlet transforming under the $Z_3$, while assuming the Majorana fermion $\chi$ transforms trivially and has a Majorana mass $m_\chi = M_1$. 
Keeping this in mind, we can map the couplings of the bino to those in the EFT, Eq.~(\ref{eq:EFTmu}), via
\begin{equation} \label{eq:pmapB}
	\delta = \alpha \to \frac{g_1^2}{2}~, \quad \lambda \to \lambda~, \quad \kappa = \xi \to 0~.
\end{equation}
The blind spot condition for the bino region is then
\begin{equation}\label{eq:bsBin}
	\sin 2\beta = - m_\chi/\mu~.
\end{equation}
\end{itemize}

\begin{figure}[tbh]
\hspace*{1mm}
\includegraphics*[width=.95\linewidth]{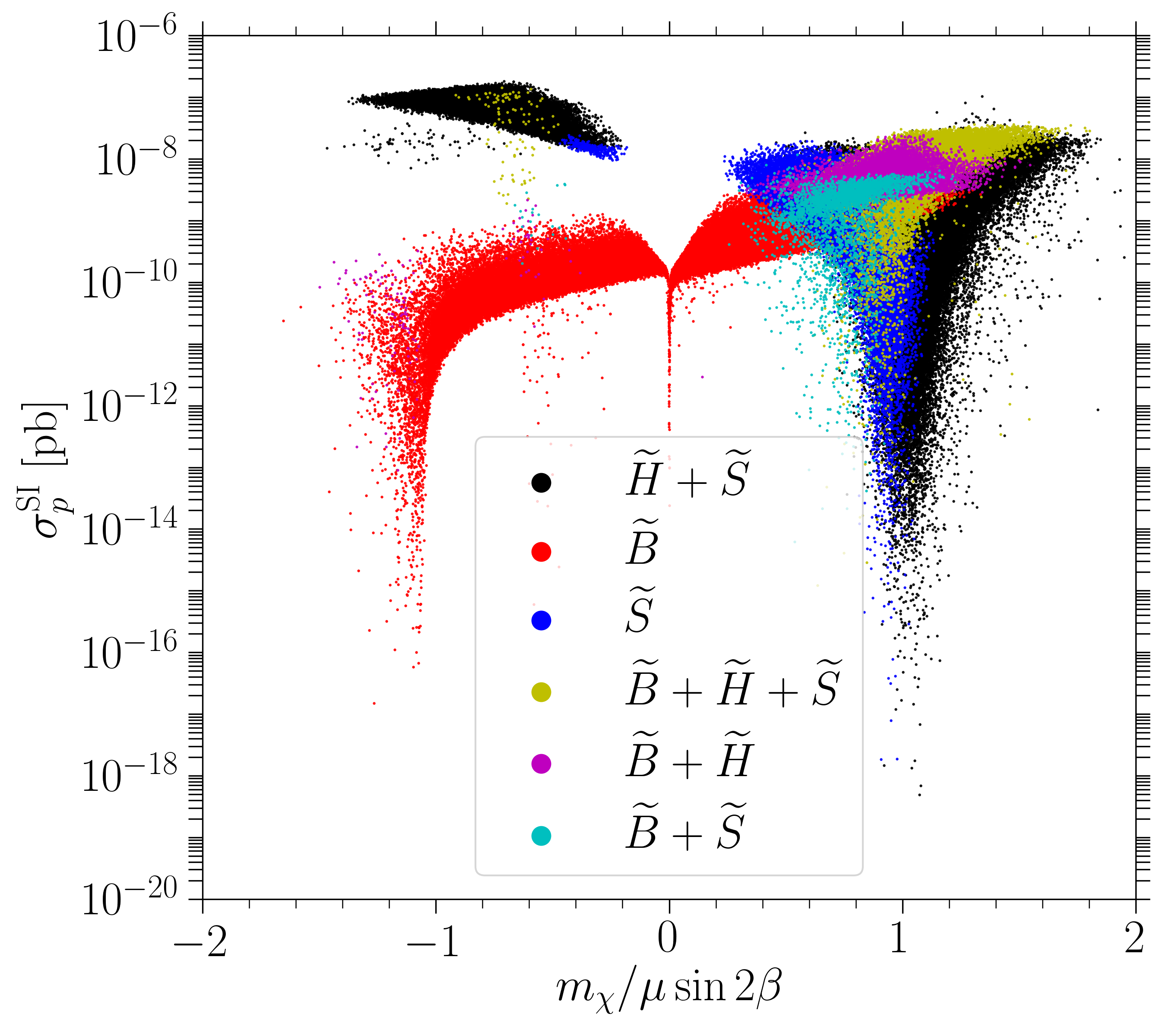} 
  \caption{\label{NMSSM_BS} 
  The SIDD cross section $\sigma_p^{\rm SI}$ f
  vs. $m_\chi/(\mu \sin 2\beta)$, where the blind-spot conditions are satisfied for $m_\chi/(\mu \sin 2\beta) = +1 (-1)$ for the singlino (bino) DM case~\cite{Baum:2017enm}.}
\end{figure}

\begin{figure}[tbh]
\hspace*{1mm}
\includegraphics*[width=.95\linewidth]{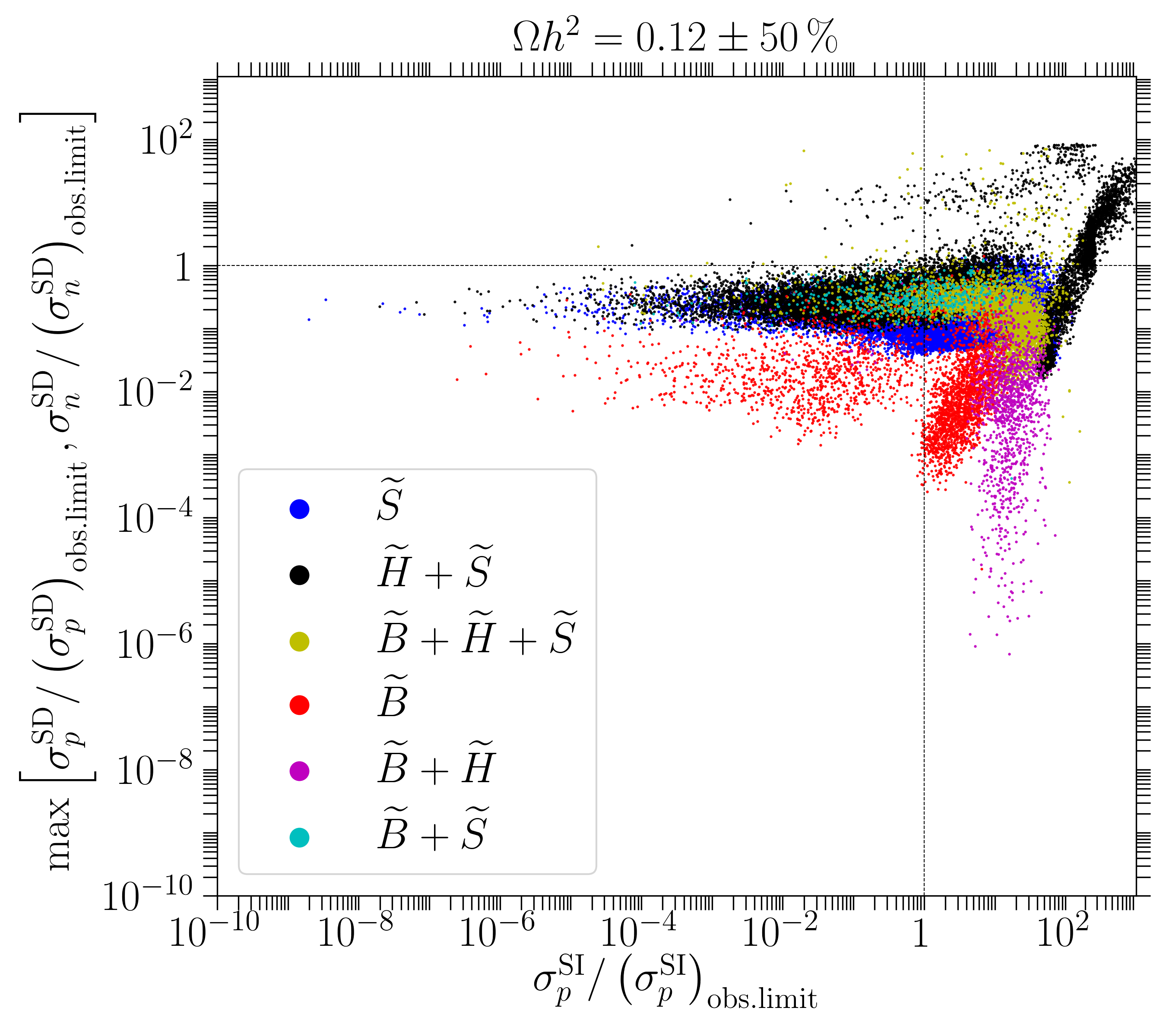} 
  \caption{\label{SISD} 
  SIDD vs. SDDD cross section in units of the respective observed limit for the same points. For SIDD cross section, at each respective DM mass we use the stronger of the two limits from XENON1T and PandaX-II. For SDDD scattering we use the more constraining of the current bounds for either SDDD scattering of neutrons from LUX~\cite{Akerib:2017kat}, or SDDD scattering of protons from PICO-60~\cite{Amole:2017dex}. To guide the eye we indicate the current bounds with thin dashed lines; points lying in the lower left quadrant satisfy all current direct detection bounds~\cite{Baum:2017enm}.}
\end{figure}

We ran numerical scan using \texttt{NMSSMTools} and \texttt{MicrOmegas} validating our expectation for the obtention of a DM with suppressed SIDD due to the presence of blindspots as dictated by Eqs.~(\ref{eq:bsSin}) and (\ref{eq:bsBin}). The results are shown in Fig.~\ref{NMSSM_BS}. We also observed that while the SIDD can be easily suppressed while obtaining the correct relic density, their is no such suppression mechanism for the spin dependent direct detection~(SDDD) cross section. In fact, while certainly the limits for SDDD are much weaker, near future prospects may allow us to probe most of the region of parameter space with very suppressed SIDD. This is shown in Fig.~\ref{SISD}, where the SDDD is seen to be at most two orders of magnitude below current limits.  

\begin{figure}[h!]
\hspace*{-1mm}
\includegraphics*[width=.95\linewidth]{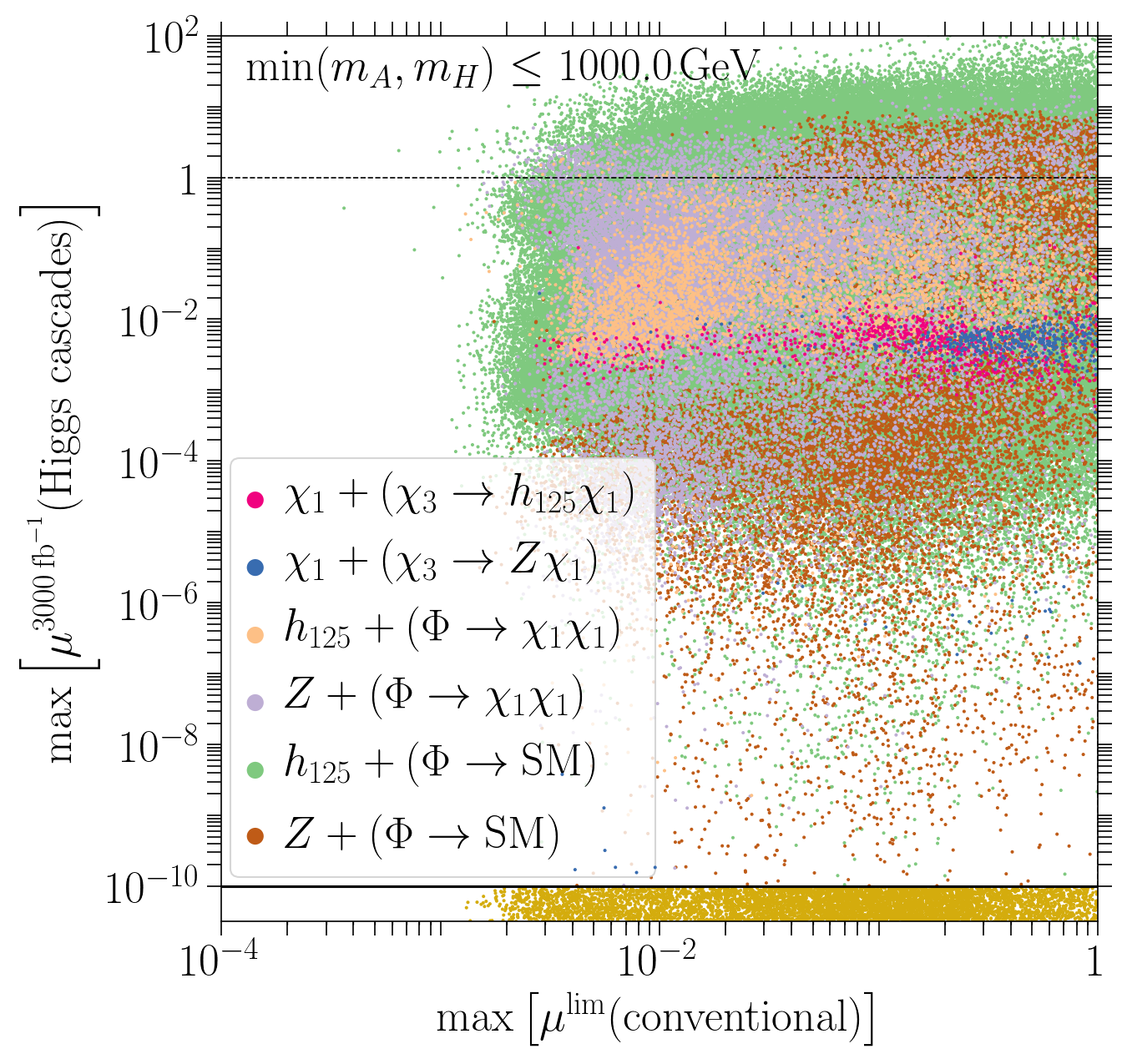} 
  \caption{\label{reach_NMSSM} 
 We present our results for the cases where at least one of the heavy Higgs bosons $H$ or $A$ lighter than 1\,TeV color coded according to the Higgs cascade channel with the largest signal strength as indicated in the legend. The $x$-axis shows the largest signal strength out of all conventional Higgs searches, 
 and the $y$-axis shows the largest signal strength out of the Higgs casccade searches. Note, that for the Higgs cascades modes we use the projected sensitivity for $L=3000\,{\rm fb}^{-1}$ of data while for the conventional searches we use the best current limit~\cite{newNMSSM}. }
\end{figure}

As mentioned earlier, the NMSSM Higgs sector is very predictive due to the presence of a SM-like $h_{125}$. We investigated the collider phenomenology associated with the Higgs sector in detail. The prospects for discovery for the scenarios with at least one mass lighter than 1 TeV are shown in Fig.~\ref{reach_NMSSM}~\cite{newNMSSM}. This shows in particular that the Higgs cascade decays channels discussed in the previous section can provide complimentary probes to the standard search channels, bringing most of the interesting parameter regions within reach of the high luminosity LHC.

\section{Summary}\label{sec:sum}
\label{sec:summ}

There has been no compelling evidence for the presence of NP at the weak scale since the Higgs discovery in 2012. This has lead to wide spread pessimism in our field regarding both the WIMP paradigm and the most popular SUSY models. In these proceedings I have presented a simple scenario where vanilla WIMP DM with thermal relic density can easily evade the stringent SIDD bounds. Using an EFT formulation I present parameter relations such that the coupling of DM can be suppressed to $h_{h125}$ while maintaining enough DM-SM interactions such that thermal equilibrium can be maintained. Given that the DM candidate is assumed to be a SM singlet, coupled with the constraints imposed by the alignment of the Higgs vacuum expectation value on an extended Higgs sector, standard search strategies at the LHC may be insensitive to NP. However, Higgs cascade channels which are unsuppressed due to the presence of the singlet scalars, may be provide an additional handle, allowing us to probe much of the relevant parameter region at the LHC.

I then present the mapping of the EFT to the NMSSM, showing that the EFT expectations are borne out using  sophisticated numerical packages available. I also show the direct detection and LHC prospects for the parameter regions discussed. 

I stress that the consistent region of parameters we obtain using both the EFT, and its mapping to the NMSSM, 
do not require any extreme choices. Indeed this region of parameters would be considered quite ``natural". That said, specific correlations between various parameters are needed. However, considering physics from the UV prospective, it may very well be that GUT scale symmetries broken near the weak scale would show up in low energy physics as strange cancellations or relationships between parameters. Nature may well have chosen to put NP at an energy scale out of our foreseeable reach, which would be our misfortune. However, it seems as likely that there may be NP at the weak scale, and we may have to be more creative to find it.



\section*{Acknowledgements}

I am grateful to the organizers
of ``FPCapri2018" and ``The Future of BSM Physics" for the
conference and workshop, and MITP for partial support
during  my stay in Capri. I also acknowledge support the U.S. Department of Energy under
Contract No. DESC0007983.


\bibliographystyle{elsarticle-num}
\bibliography{mybib}







\end{document}